\documentclass[aps,twocolumn,nofootinbib]{revtex4-1}
\usepackage{epsfig}
\usepackage{graphicx}
\usepackage{amsmath,amssymb,amsfonts}
\usepackage{array}
\usepackage{url}
\usepackage{multirow}
\usepackage{natbib}
\usepackage{float}
\usepackage{lineno}
\usepackage{xspace}
\usepackage{ulem}
\usepackage{footnote}

\usepackage[usenames,dvipsnames]{color}
\definecolor{darkblue}{RGB}{0,0,196}
\usepackage[colorlinks=true,linkcolor=darkblue,citecolor=darkblue,urlcolor=darkblue]{hyperref}

\begin{document}
\title{Energy flow in ultra-high energy cosmic ray  interactions as a probe of thermalization: a potential solution  to the muon puzzle}

\author{Ronald Scaria$^1$}
\email[]{ronaldscaria.rony@gmail.com}
\author{Suman Deb$^2$}
\email[]{sumandeb0101@gmail.com}
\author{Captain R. Singh$^1$}
\email[]{captainriturajsingh@gmail.com}
\author{Raghunath Sahoo$^1$}
\email[Corresponding author:]{Raghunath.Sahoo@cern.ch}
\affiliation{$^1$Department of Physics, Indian Institute of Technology Indore, Simrol, Indore 453552, India}
\affiliation{$^2$Laboratoire de Physique des 2 infinis Irène Joliot-Curie, Université Paris-Saclay, CNRS-IN2P3, F-91405 Orsay, France}

\begin{abstract}

Signatures of the formation of a strongly interacting thermalized matter of partons have been observed in nucleus-nucleus, proton-nucleus, and high-multiplicity proton-proton collisions at LHC energies. Strangeness enhancement in such ultra-relativistic heavy-ion collisions is considered to be a consequence of this thermalized phase, known as quark-gluon plasma (QGP). Simultaneously, proper modeling of hadronic energy fraction in interactions of ultra-high energy cosmic rays (UHECR) has been proposed as a solution for the ``muon puzzle”, an unexpected excess of muons in air showers. These interactions have center-of-mass collision energies of the order of energies attained at the LHC or even higher, indicating that the possibility of a thermalized partonic state cannot be overlooked in UHECR-air interactions. This work investigates the hadronic energy fraction and strangeness enhancement to explore QGP-like phenomena in UHECR-air interactions using various high-energy hadronic models. A core-corona system with a thermalized core undergoing statistical hadronization is considered through the EPOS LHC model. In contrast, PYTHIA 8, QGSJET II-04, and SYBILL 2.3d consider string fragmentation without thermalization. We have found that EPOS LHC gives a better description of strangeness enhancement as compared to other models. We conclude that adequately treating all the relevant effects and further retuning the models is necessary to explain the observed effects.

\end{abstract}
\date{\today}
\maketitle

\section{Introduction}
\label{into}
Ultra-High Energy Cosmic Ray (UHECR) experiments provide the ideal opportunity to study particle physics at center-of-mass energies and kinematic regions, which are inaccessible at accelerator facilities like the Large Hadron Collider (LHC)~\cite{Anchordoqui:2018qom}. The primary UHECRs interact with atomic nuclei in the atmosphere, producing multitudes of secondary particles which further interact or decay depending upon their energy. This creates a shower of particles spreading over a vast area, termed as extensive air showers (EAS). These EAS are measured on the ground to understand the nature and the origin of the primary cosmic rays. Determining cosmic ray mass composition as a function of primary energy is also interesting for ground-based EAS experiments. Such measurements are based mainly on two shower features: (a) the depth of shower maximum, $X_{max}$, and (b) the number of muons produced, $N_\mu$~\cite{Albrecht:2021cxw,Ostapchenko:2016dxc}. However, precise determination of the mass composition is limited by different model predictions of these features, which are based on extrapolations to the hadronic interaction models tuned to explain collider data.\\

The Pierre Auger observatory \cite{PierreAuger:2014ucz,PierreAuger:2016nfk} and the Telescope Array~\cite{TelescopeArray:2018eph} observe that various models consistently underestimate the number of muons in an air shower. Different experiments have carried out similar measurements, prompting a system-independent review of muon measurements \cite{EAS-MSU:2019kmv}. The results indicate a consistent muon excess in multiple experiments over a wide energy range, starting smoothly and increasing logarithmically in primary energy above the ``knee"\footnote{``Knee" refers to the region ($\sim 4$ PeV) of the cosmic ray (CR) energy spectrum where an abrupt change in the spectral index is observed. This is usually attributed to the gradual change of the CR source from galactic to extragalactic origin \cite{IceCube:2013ftu}.} of the cosmic ray energy spectra. The Pierre Auger observatory has also explored the energy dependence of this muon excess at very high primary energies~\cite{PierreAuger:2014ucz}. Moreover, a novel study on the shower-to-shower fluctuations of $N_\mu$ ~\cite{PierreAuger:2021qsd} suggests that this muon deficit in models might be a small deficit at each step that gets accumulated over the shower development. At the same time, the measurement of $X_{max}$ agrees fairly well with model estimates. Parameter tuning of various models has been unsuccessful in solving this muon discrepancy which is now widely termed as the ``muon puzzle".

Lattice QCD (lQCD) calculations \cite{Borsanyi:2010cj} predict a transition of ordinary nuclear matter to a deconfined state of partons known as the Quak-Gluon Plasma (QGP) under extreme energy densities and temperature. The ultra-relativistic collisions at Relativistic Heavy Ion Collider (RHIC) and the LHC have achieved such extreme conditions that favor the formation of a phase of thermalized QCD matter.
A simple calculation yields that the energy deposited per unit nuclear overlap area at the time of collision in proton-air, helium-air, and nitrogen-air collisions are higher than or comparable with Pb-Pb energy densities at the LHC~\cite{Anchordoqui:2019laz}. Therefore the formation of such a deconfined phase cannot be ruled out in interactions of high-energy cosmic rays with air nuclei. Strangeness enhancement in the final state has been used in relativistic nuclear collision experiments as a signature of QGP~\cite{Koch:2017pda,Rafelski:1982pu}. The ALICE collaboration has recently reported the observation of strangeness enhancement even in small systems~\cite{ALICE:2016fzo}. It has also been observed in $pp$ collisions using zero degree calorimeter (ZDC) that strangeness enhancement is inversely related to the energy deposited in the ZDC~\cite{Schotter:2023khz}. This suggests that the medium formation possibility depends highly on the energy deposited at the collision vertex. Thus, the energy available for particle production and further shower development
is important in UHECR-air interactions. One could also, in principle, explore the effect of this showering by studying strange particle yield at different distances from the initial interaction vertex using the CORSIKA air shower simulation package \cite{Heck:1998vt}. Such a study ~\cite{Scaria:2022ugk} indicates that the strangeness component increases with primary energy and with decreasing distance from the interaction vertex. Considering a simple model where the initial energy is distributed equally among the daughter hadrons \cite{Matthews:2005sd,Ulrich:2010rg}, an increase of strange particle yield would mean that the energy transferred from the hadronic cascade to electromagnetic cascade through $\pi^0\xrightarrow{} 2\gamma$ decay at each step is reduced. This would, in turn, increase the number of muons at the ground level due to meson decays. The formation of QGP in cosmic ray interactions in the atmosphere has thus been cited as a possible solution to the muon puzzle~\cite{Albrecht:2021cxw,Anchordoqui:2019laz,Petrukhin:2014lma}. \\

The amount of energy available for hadron production and subsequent shower development is an important parameter that drives the muon multiplicity of an air shower ~\cite{Ulrich:2010rg,Pierog:2006qv,Cazon:2018gww,Baur:2019cpv}. Recently, $p$-O and O-O collisions have been proposed at the LHC, specifically emphasizing cosmic ray-related measurements~\cite{Brewer:2021kiv}. It would thus be fruitful to look at the energy division between electromagnetic and hadronic particles in such collisions. This may be quantified by ~\cite{Perlin:2021rwh,Baur:2019cpv},
\begin{equation}
\label{eq1}
    R(\eta) = \frac{\langle dE_{em}/d\eta \rangle}{\langle dE_{had}/d\eta \rangle}
\end{equation}
where $\langle dE_{em}/d\eta \rangle$ is the average energy carried by photons and $e^\pm$ while $\langle dE_{had}/d\eta \rangle$ is the average energy summed over all hadrons in bins of pseudorapidity, $\eta$. This quantity is related to the hadronization mechanism that the partonic system follows. A system with high energy density is expected to follow statistical hadronization, favoring the production of heavier hadrons. Consequently, more charged hadrons are produced as compared to $\pi^0$ mesons. This, in turn, reduces the energy lost to the electromagnetic cascade.\\

As observed at the LHC, the possibility of the formation of QGP in small systems~\cite{ALICE:2016fzo,CMS:2010ifv,CMS:2016fnw,PHENIX:2017djs} piques one's interest in the possibility of such a medium formation in cosmic ray interactions. We aim to explore various hadronization schemes used in different high-energy cosmic ray models by studying strangeness and the energy fraction $R$ and exploring the connection between these terms. 
Section \ref{sec2} details the theoretical models and the chosen phase space kinematics. The results exploring strangeness and $R$ are shown in sec.\ref{Sec3} along with their discussion before finally summarizing the results in sec.\ref{sum}.


\section{Brief Description of models considered}
\label{sec2}

\begin{table*}[htp]
\caption{The colliding species considered, the center of mass energies, and the models used in the present study. The models chosen are EPOS LHC~\cite{Pierog:2013ria}, QGSJET II-04~\cite{Ostapchenko:2010vb}, SYBILL 2.3d~\cite{Riehn:2019jet}, PYTHIA 8~\cite{Bierlich:2022pfr,Bierlich:2018xfw}. The color reconnection tune with gluon splitting is chosen for $pp$, while ANGANTYR mode with rope hadronization and color reconnection are chosen for Pb-Pb within PYTHIA 8.}
\vspace{0.5cm}

\centering 
\scalebox{1.25}
{
\begin{tabular}{|c|c|c|c|}
\hline  
    \multicolumn{1}{|c|}{System}&\multicolumn{1}{c|}{Colliding Energy (TeV)}&\multicolumn{1}{c|}{Models} \\
\cline{2-3}

\hline

\multirow{1}{*}{$pp$}  

& 7 \& 13 & EPOS LHC, QGSJET II-04, SYBILL 2.3d and PYTHIA 8 \\
\cline{2-3} 
\cline{1-3} 

\multirow{1}{*}{$p$-O} 

& 9.9 & EPOS LHC \\
\cline{2-3} 
\cline{1-3} 

\multirow{1}{*}{$p$-Pb} 

& 5.02 \& 8.16 & EPOS LHC \\
\cline{2-3} 
\cline{1-3} 

 \multirow{1}{*}{Pb-Pb} 

& 2.76 \& 5.02 & EPOS LHC and PYTHIA ANGANTYR \\

\cline{2-3} 
\cline{1-3} 

 \end{tabular}
}
\label{tab:mult_sp}
\end{table*}

Particle production in ultra-relativistic heavy-ion collisions is treated using perturbative and/or non-perturbative quantum chromodynamic (QCD) methods. There are many models considering various QCD processes for particle production, which are quite successful in explaining some of the experimental data. In this work, we chose four models, three of which are updated versions of frequently used cosmic ray high-energy interaction models. The fourth finds more application in accelerators. These models are being tested using various collision species and different center-of-mass collision energies, the details of which are given in Table ~\ref{tab:mult_sp}.
The EPOS LHC~\cite{Pierog:2013ria}, QGSJET II-04~\cite{Ostapchenko:2010vb} and SYBILL 2.3d~\cite{Riehn:2019jet} models are provided within the Cosmic Ray Monte Carlo Package, CRMC (v2.0.1)~\cite{CRMC} while for the PYTHIA 8~\cite{Bierlich:2022pfr,Bierlich:2018xfw} tunes, we have used the PYTHIA 8305 version.

Of the cosmic ray interaction models, EPOS LHC and QGSJET II-04 are based on the ``semihard Pomeron" approach~\cite{Drescher:1999zy,Ostapchenko:2001hz,Drescher:2000ha} within the Reggeon Field Theory (RFT) ~\cite{Gribov1968}. This allows the inclusion of both ``soft" and ``hard" processes in the interaction mechanism by the introduction of a ``soft-hard" separation scale~\cite{Ostapchenko:2016dxc}. This effectively divides the evolution into perturbative and non-perturbative regimes. The perturbative part is described by Dokshitzer-Gribov-Lipatov-Altarelli-Parisi (DGLAP) formalism~\cite{Gribov:1972ri,Dokshitzer:1977sg,Altarelli:1977zs}, while the non-perturbative soft part is described as soft Pomeron emissions. SYBILL 2.3d, on the other hand, uses a Dual Parton Model and the minijet model where an eikonal approximation is used in the impact parameter space to determine the total scattering amplitudes, which in turn decide the interaction cross sections. The hadronization scheme of EPOS LHC uses a core-corona approach where the core hadronizes statistically while the corona uses Lund string fragmentation~\cite{Pierog:2013ria,Andersson:1983jt}. This differs from the QGSJET II-04 and SYBILL 2.3d models, which use Lund string fragmentation only~\cite{Ostapchenko:2010vb,Riehn:2019jet}. In Ref.~\cite{Baur:2019cpv}, the authors have tried to explore the effect of the hadronization scheme in explaining the muon data of Pierre Auger experiment ~\cite{PierreAuger:2014ucz} by changing the core contribution in both EPOS LHC and QGSJET II-04 models.

PYTHIA, on the other hand, is a pQCD-inspired event generator that has successfully explained many results at the LHC~\cite{Bierlich:2022pfr}. The event develops through hard and soft scatterings, which leads to initial and final parton showers. One parton may interact with multiple other partons using the multi-parton interaction option. Hadronization is mainly based on the Lund string fragmentation method, which may be aided through the color reconnection scheme or replaced with the rope hadronization mechanism. Finally, the particles can undergo rescattering, regeneration, and other final state effects. The recent development of the ANGANTYR~\cite{Bierlich:2018xfw} model allows one to carry out heavy ion (A-A and $p$-A) collisions using the underlying physics of PYTHIA.

One million events each are generated for $pp$, $p$-O, and $p$-Pb collisions, while for the Pb-Pb system, 500 thousand events are generated at the corresponding energies. We have chosen the pseudorapidity range $|\eta|<2.0$ in our calculations for all systems~\cite{ALICE:2017pcy}. The $c\tau$ definition used at the ALICE experiment~\cite{ALICE:2017pcy} defines the final state particles. These particles are further used to measure the $R$ factor and strangeness ($K/\pi$) in the collision as a function of charged particle multiplicity.

\section{Results and Discussion}
\label{Sec3}

\begin{figure*}[!ht]
\includegraphics[scale=0.4]{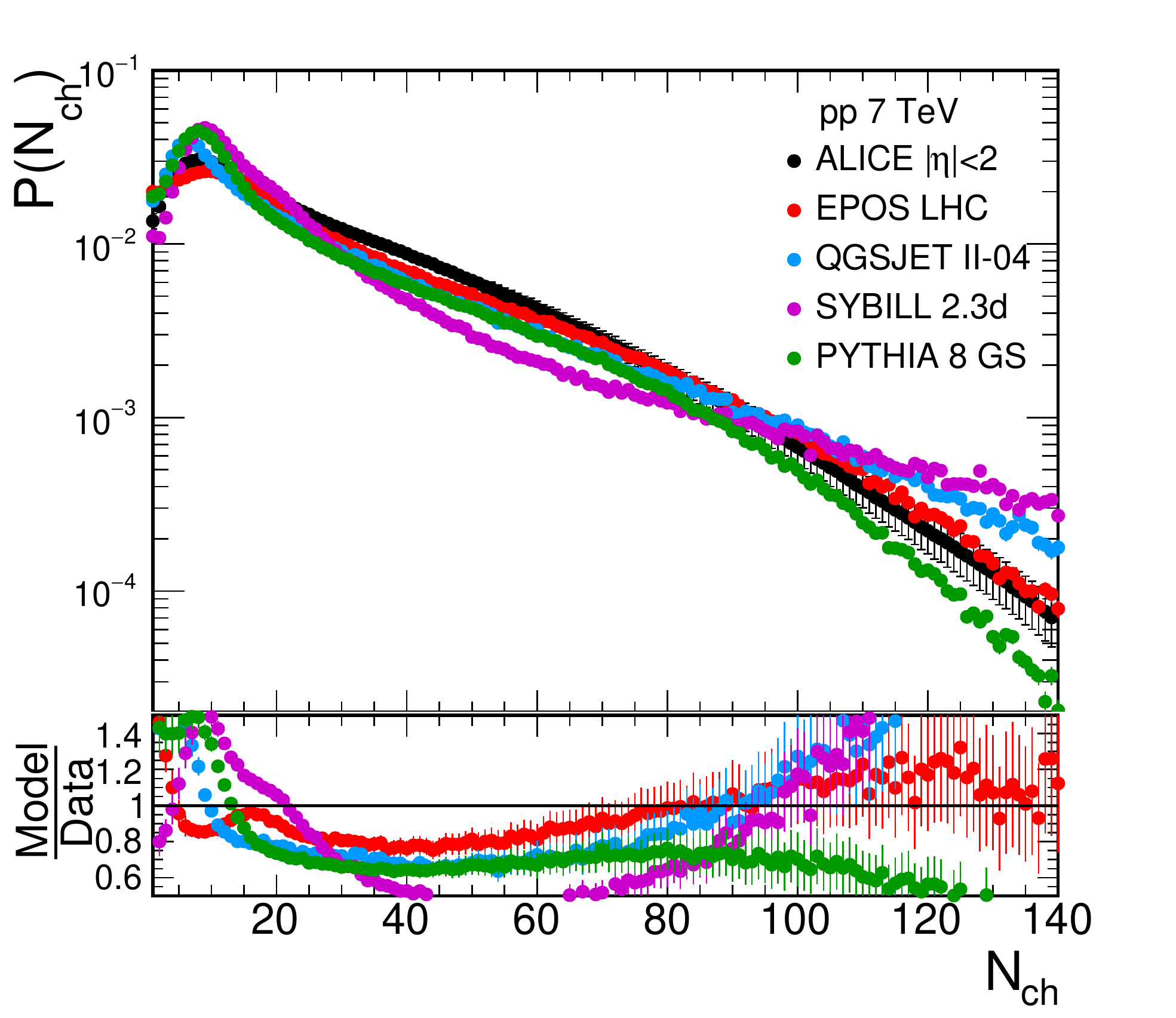}
\includegraphics[scale=0.4]{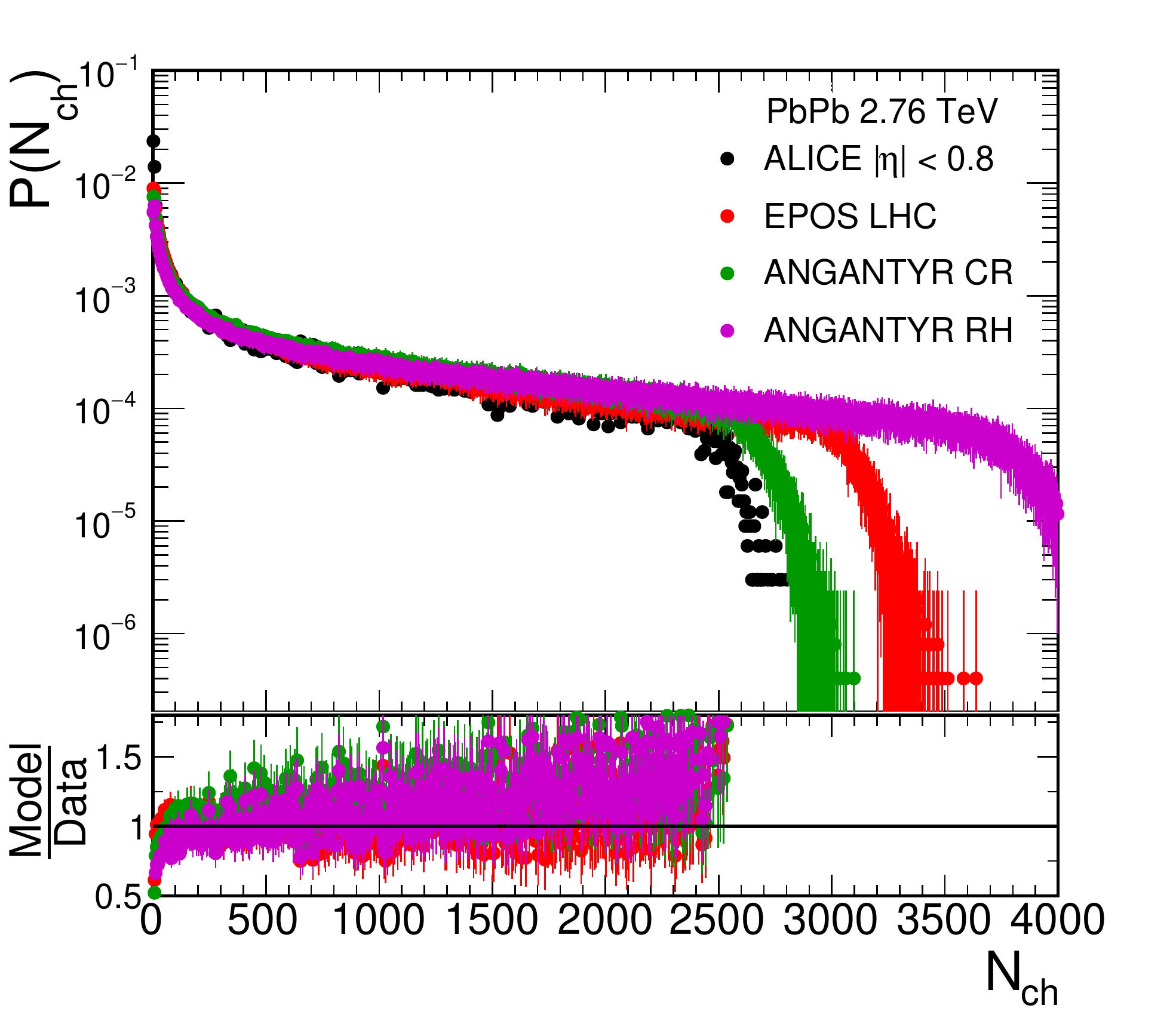}
\caption {(Color online) Upper left panel: Charged-particle multiplicity distribution for $pp$ collisions at $\sqrt{s} =$ 7 TeV obtained from various models and ALICE experimental data ~\cite{ALICE:2017pcy} Lower left panel: Ratio of simulated data obtained from various models and experimental data~\cite{ALICE:2017pcy}. Upper right panel: Charged-particle multiplicity distribution for Pb-Pb collisions at $\sqrt{s} =$ 2.76 TeV obtained from various models and ALICE experimental data ~\cite{ALICE:2010suc} Lower right panel: Ratio of simulated data obtained from various models and experimental data~\cite{ALICE:2010suc}. The error bars in the data points are the statistical uncertainties.}
\label{fig1}  
\end{figure*}

To ensure the compatibility and quality of the data sample used for the present work, we have compared the simulated data obtained from the EPOS LHC, QGSJET II-04, SYBILL 2.3d, PYTHIA 8, and PYTHIA ANGANTYR with the ALICE experimental data~\cite{ALICE:2017pcy,ALICE:2010suc} in the same kinematic range. Fig.~\ref{fig1} shows this comparison of charged-particle multiplicity distribution for $pp$ and Pb-Pb collisions at $\sqrt{s} =$ 7 and 2.76 TeV, respectively, between mentioned models and ALICE data. For the $pp$ case, all the models give a reasonable description of the P($N_{ch}$) distribution over two-orders of magnitude within $\pm 50$\%. For the Pb-Pb case, the shape of the $N_{ch}$ distributions are similar in all models, but the multiplicity in the most central events is overpredicted by 25\% in EPOS LHC and around 50\% in PYTHIA ANGANTYR with rope hadronization (RH). From the lower panels of  Fig~\ref{fig1}, it is evident that the EPOS LHC (PYTHIA ANGANTYR with color reconnection (CR)) are better in agreement with the experimental results for the $pp$ (Pb-Pb) system.\\

With this ``quality assurance" study, we divide the data into ten equal multiplicity (centrality) classes. For simplicity, we have considered (0-10)\% as the most central or high-multiplicity events, (90-100)\% as the peripheral or low multiplicity events, and minimum biased (0-100)\% as the multiplicity integrated events for the rest of the analysis. 


\subsection{Strangeness Production through various observables and their correlation}
\label{subsec1}

\begin{figure}[ht!]
\includegraphics[scale = 0.4]{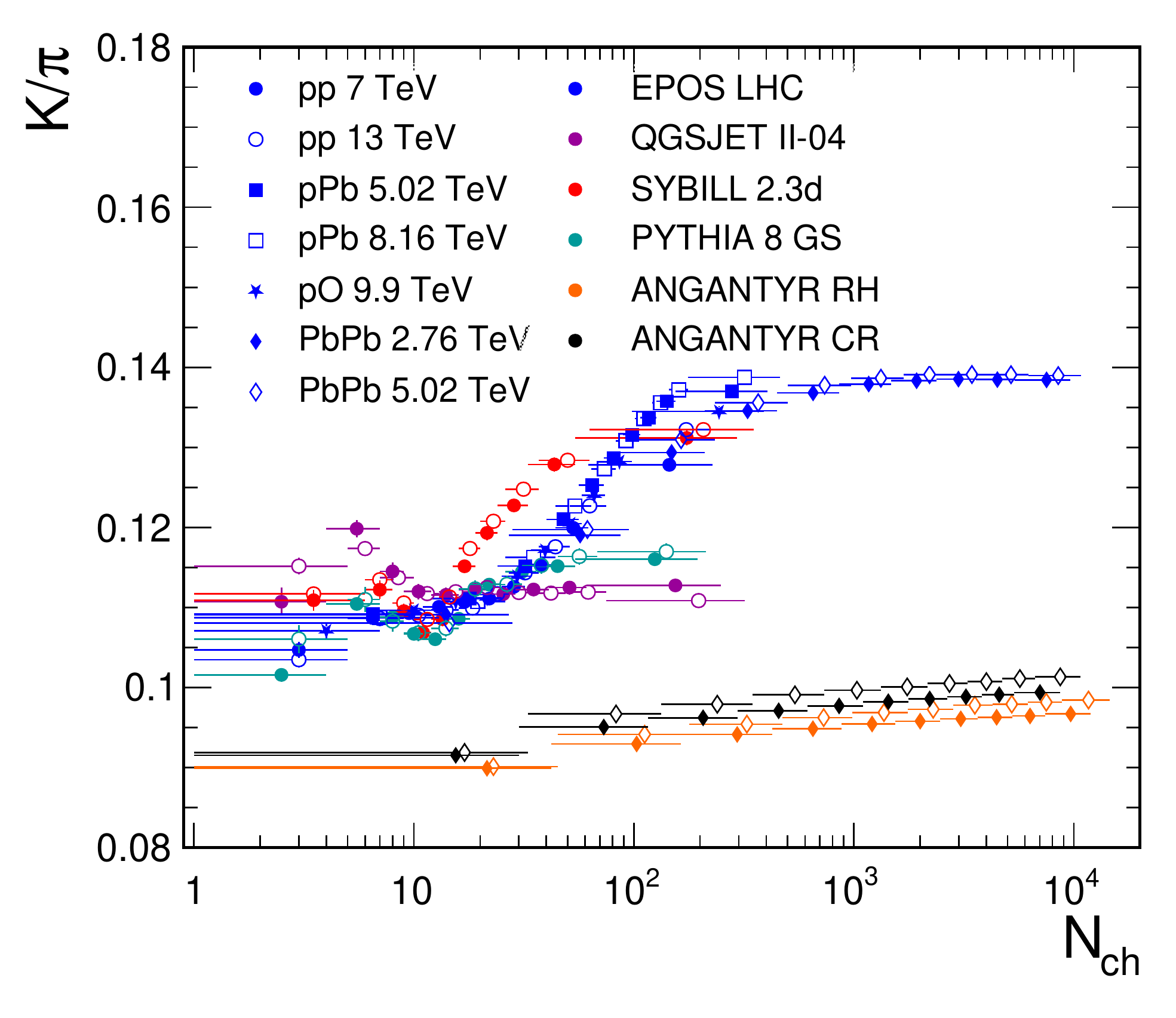}
\caption{(Color Online) Ratio of kaon to pion meson yields with mean charged particle multiplicity across various colliding species and center of mass energies considering simulated data from multiple models. Here the symbols denote different collision species and the colors represent different theoretical models as mentioned in the figure legend.}
\label{fig2}
\end{figure}

The strangeness production mechanisms of various models are studied through the evolution of kaons to pions ($\rm{K/\pi}$) ratio with charged-particle multiplicity. Fig~\ref{fig2} shows this comparison. Here, $\rm{K/\pi}$ ratio is studied across various colliding species from $pp$ to Pb-Pb through $p$-O and $p$-Pb to have a sense of system size effect and across the existing/proposed center of mass energies. Following experimental results~\cite{ALICE:2020nkc}, this ratio is expected to rise with an increase in multiplicity followed by saturation towards a very large system size (See Appendix \ref{app_A}). It is observed that EPOS LHC and SYBILL 2.3d follow this rising trend. However, SYBILL 2.3d seems to have a dip in the region $7< \rm{N}_{ch}<11$, which is in contradiction to the present experimental results~\cite{ALICE:2020nkc}. Compared to all the models considered in the present work, EPOS LHC seems to work in line with the experimental prediction~\cite{ALICE:2020nkc}. This observation points towards the importance of the core-corona treatment used in EPOS LHC to account for the strangeness enhancement 
even in small systems like $pp$ and the goodness of the model to carry out further analysis.\\

\begin{figure}[ht!]
\includegraphics[scale = 0.4]{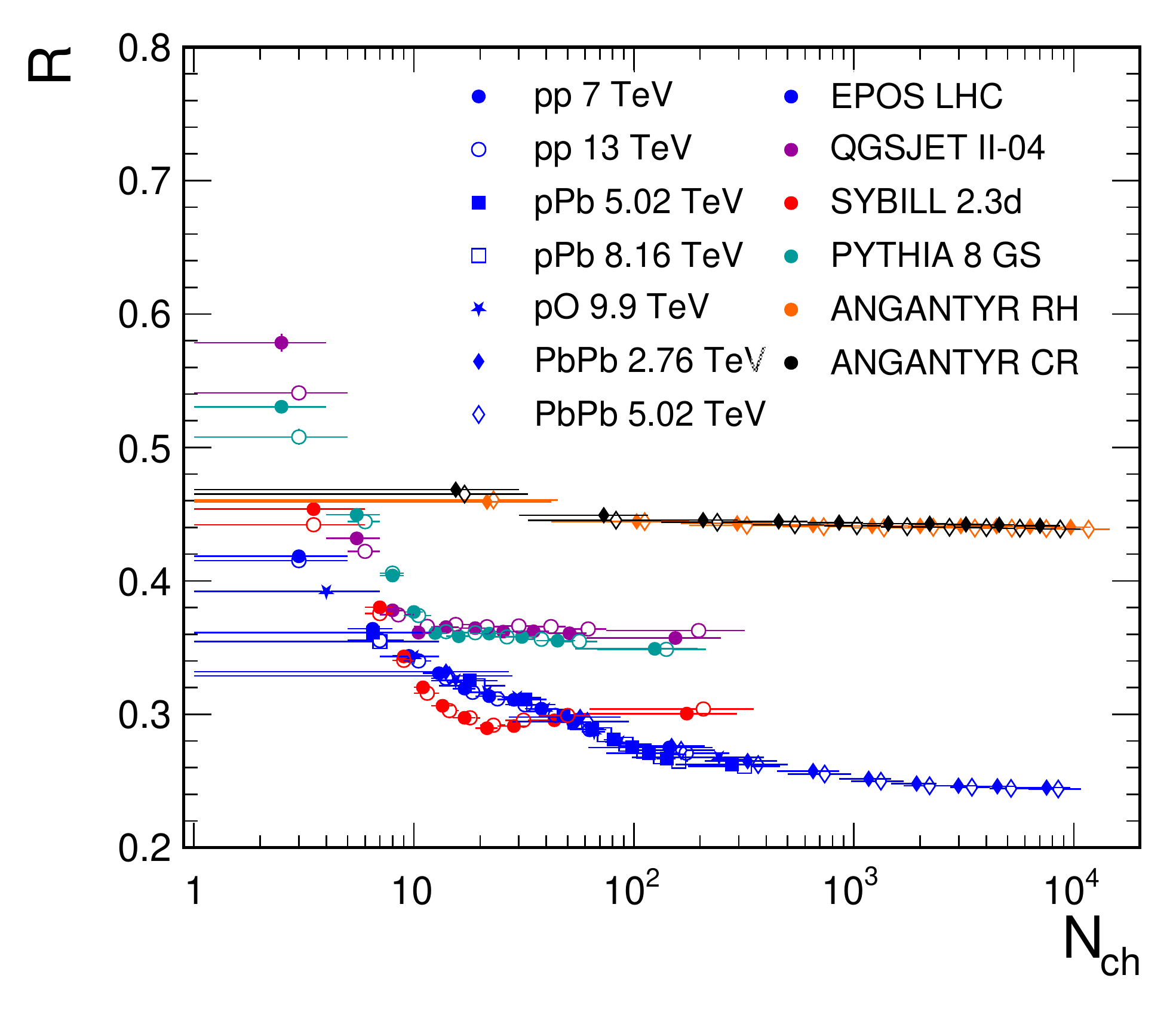}
\caption{(Color Online) Variation of the observable $R$ as a function of charged particle multiplicity across various colliding species and center of mass energies considering simulated data from multiple models. Here the symbols denote different collision species and the colors represent different theoretical models as mentioned in the figure legend.}
\label{fig3}
\end{figure}

As stated in the introduction, energy densities of various primary collisions in UHECR-air interactions could be comparable to heavy ion collisions (HIC) at the LHC. 
Air showers develop through multiple interactions spread over large distances from the initial hard interaction. Thus, commonly used observables of thermalized medium formation at colliders become redundant in the case of UHECR-air interactions.
Further, the shower development is mainly driven by the leading hadron produced in a collision~\cite{Anchordoqui:2016oxy}. The leading hadron is the hadron that carries away maximum energy from the interaction vertex. Due to their large energies, they are more likely to collide further with air nuclei, thus developing the air shower. Hence, proper modeling of the energy fraction $R$ proposed in Eq.~\ref{eq1} in different models is imperative for correctly describing UHECR-air interactions. As can be deduced from Eq.~\ref{eq1}, an increase in strange particle production would lead to a decrease in the value of $R$, which may thus be considered as the consequence of the formation of a thermalized medium. Fig.~\ref{fig3} shows $R$ variation as a function of charged particle multiplicity across various colliding species and center of mass energies considering simulated data from EPOS LHC, QGSJET II-04, SYBILL 2.3d, PYTHIA  8 and PYTHIA ANGANTYR models. It is observed that the value of $R$ for the EPOS LHC decreases with an increase in charged particle multiplicity. For the other models, it remains relatively constant (QGSJET II-04, PYTHIA 8, and PYTHIA ANGANTYR) or decreases and then increases (SYBILL 2.3d). This observation aligns with Fig.~\ref{fig2} and points towards the possibility of enhancement in strange hadron production. Interestingly, such a decreasing trend of $R$ is observed in the $p$-O system, which is the relevant interaction species for EAS development.\\

\begin{figure}[ht!]
\includegraphics[scale = 0.4]{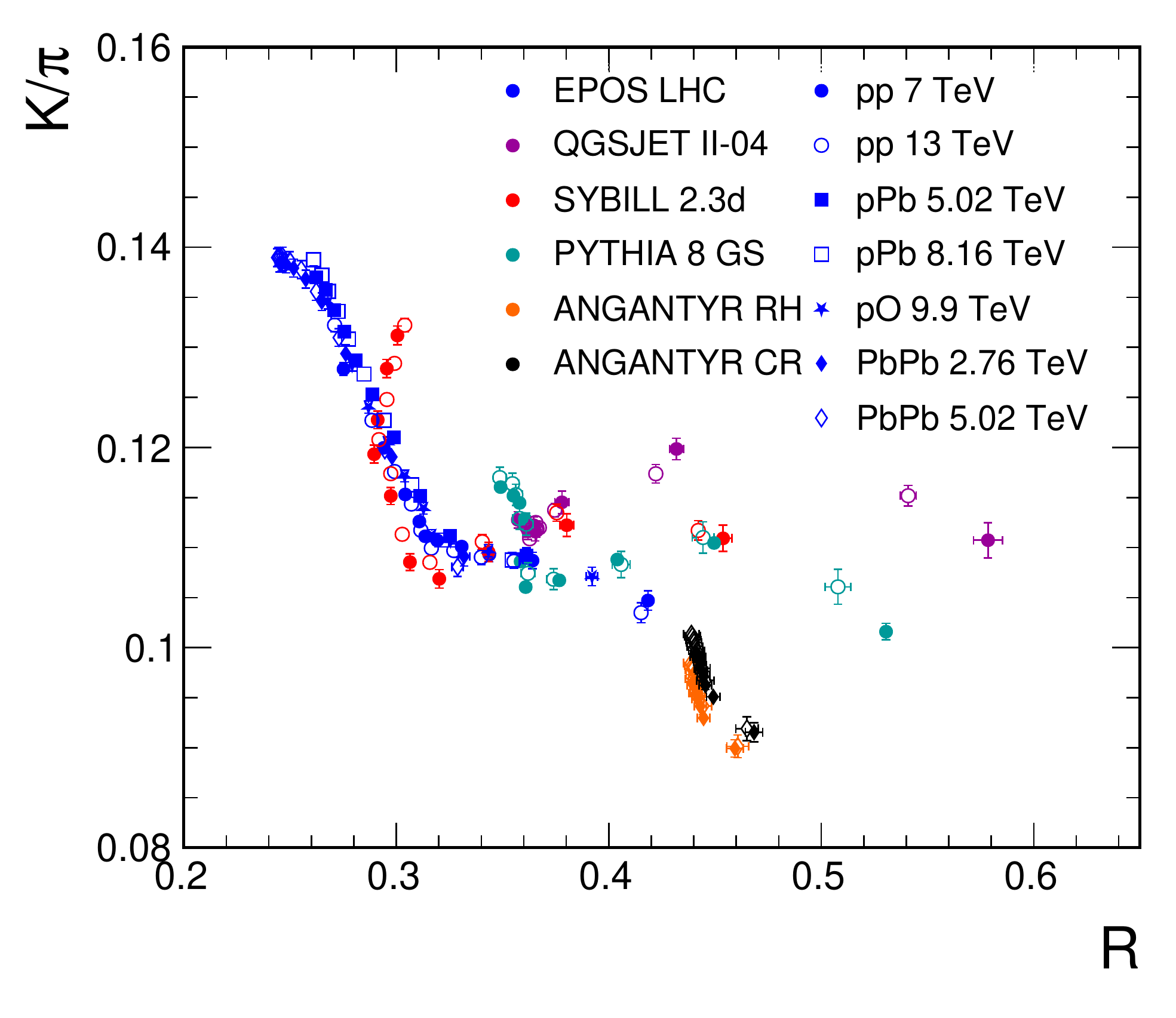}
 \caption{(Color online) Correlation between $\rm{K/\pi}$ ratio and $R$ using different models shown over various colliding species and corresponding collision energies.}
\label{fig4}
\end{figure}

From Fig.~\ref{fig2} and \ref{fig3}, it is also observed that the variation of $\rm{K/\pi}$ ratio and $R$ depends on the multiplicity. Hence, studying the correlation between these observables with multiplicity is imperative. This analysis would point towards the relevance of a threshold multiplicity (similar to the results in \cite{Thakur:2017kpv,Hatwar:2022cbx,Sahu:2020nbu,Sahu:2020swd}) in relation to an investigation of the observable $R$ in collider experiments like LHC. With this motivation, we have made a correlation study between $\rm{K/\pi}$ ratio and $R$ as shown in Fig.~\ref{fig4}. Here, we observe a negative linear correlation for EPOS LHC simulated data for the range $N_{ch} \approx 10 - 200 $, and the linear correlation breaks beyond these limits. This correlation is expected as a rise in the $\rm{K/\pi}$ ratio at LHC energies implies a decrease in the observable $R$. However, it is interesting that the correlation holds in the relevant region for systems like $pp$, $p$-O, and $p$-Pb. All other models seem to violate such a correlation drastically. This result suggests that proper modeling of the energy fraction $R$ is essential for exploring the possibility of thermalized medium formation in cosmic showers and solving the muon puzzle.

\begin{figure}
\includegraphics[scale = 0.4]{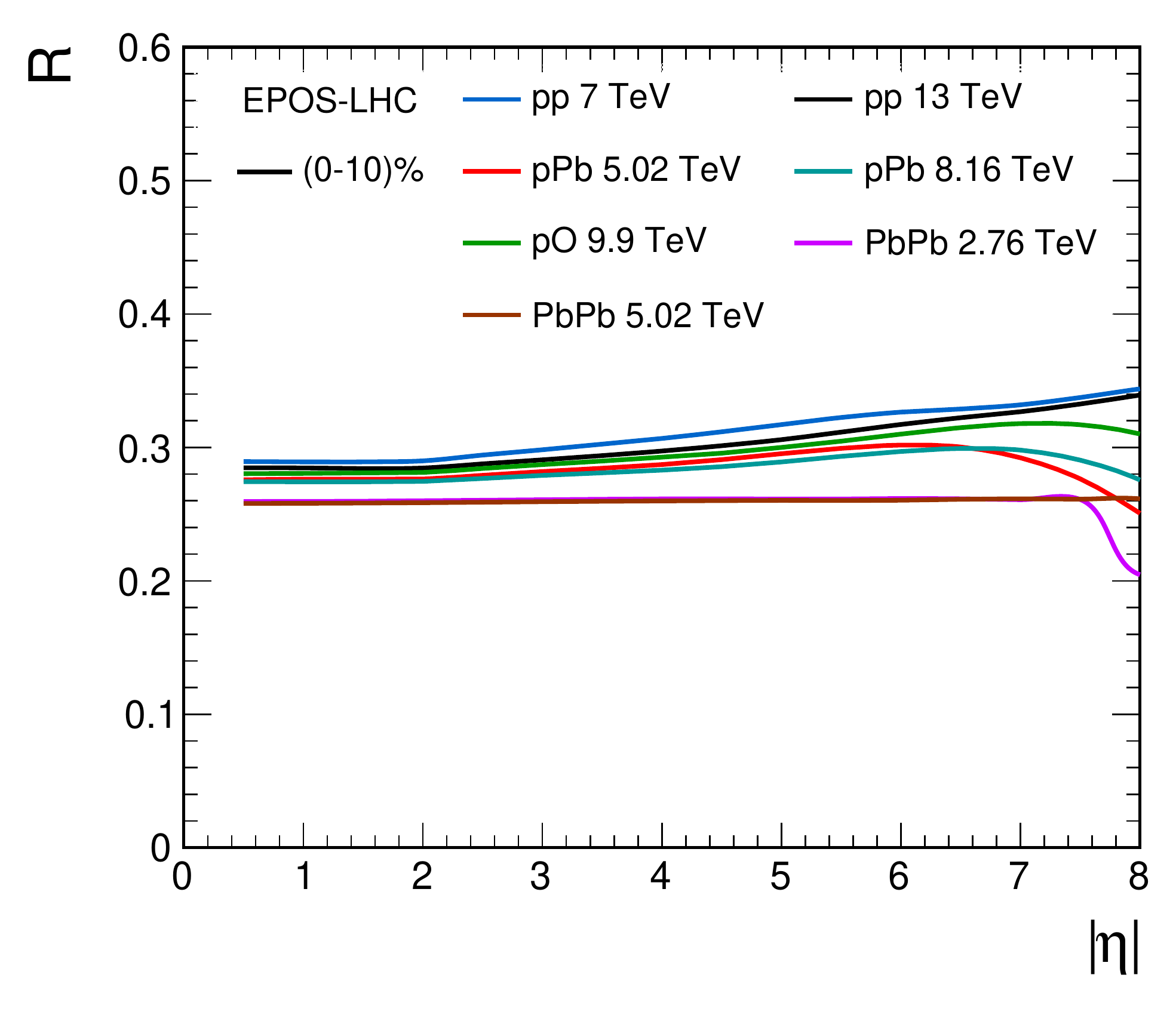}
\includegraphics[scale = 0.4]{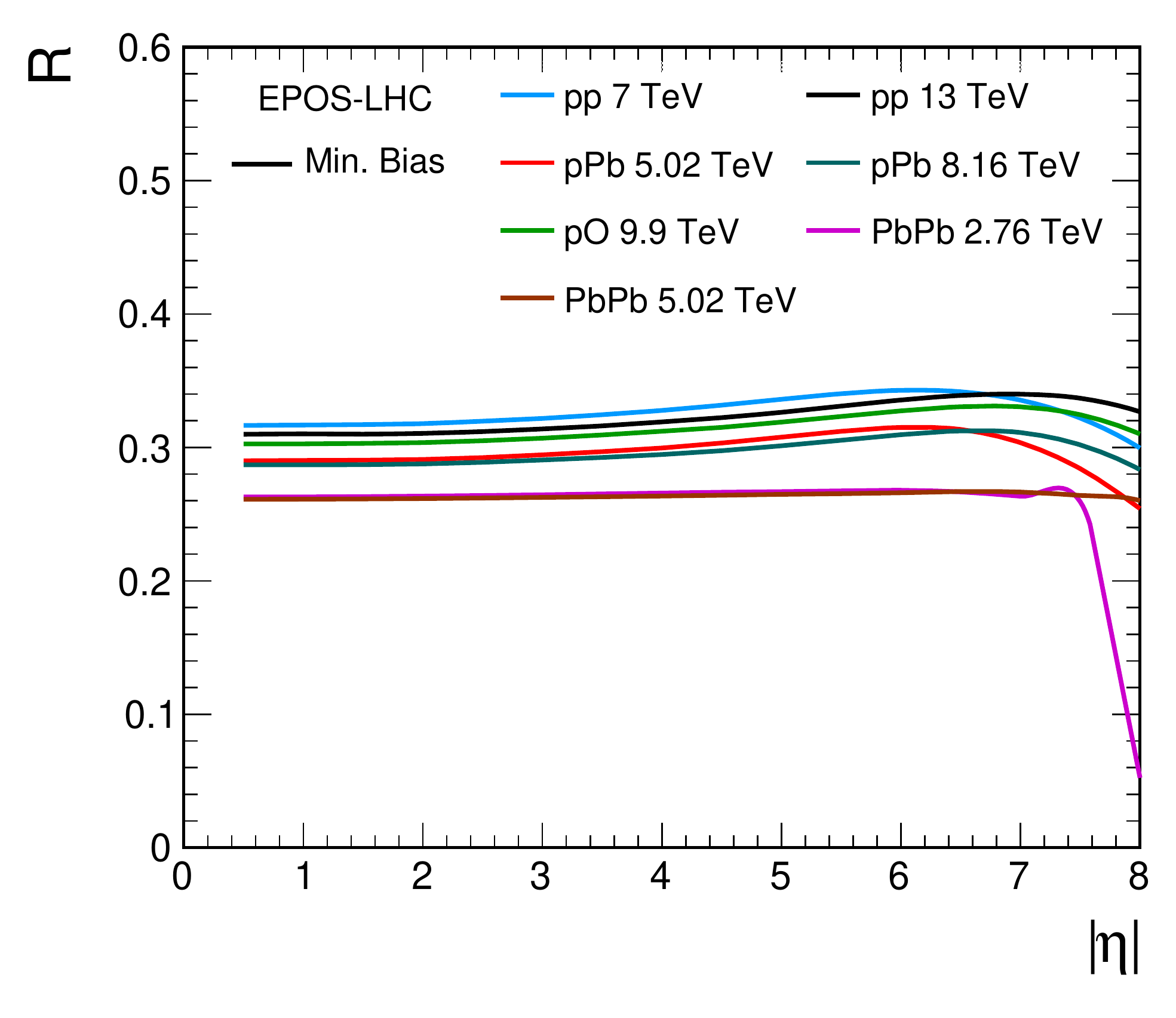}
\includegraphics[scale = 0.4]{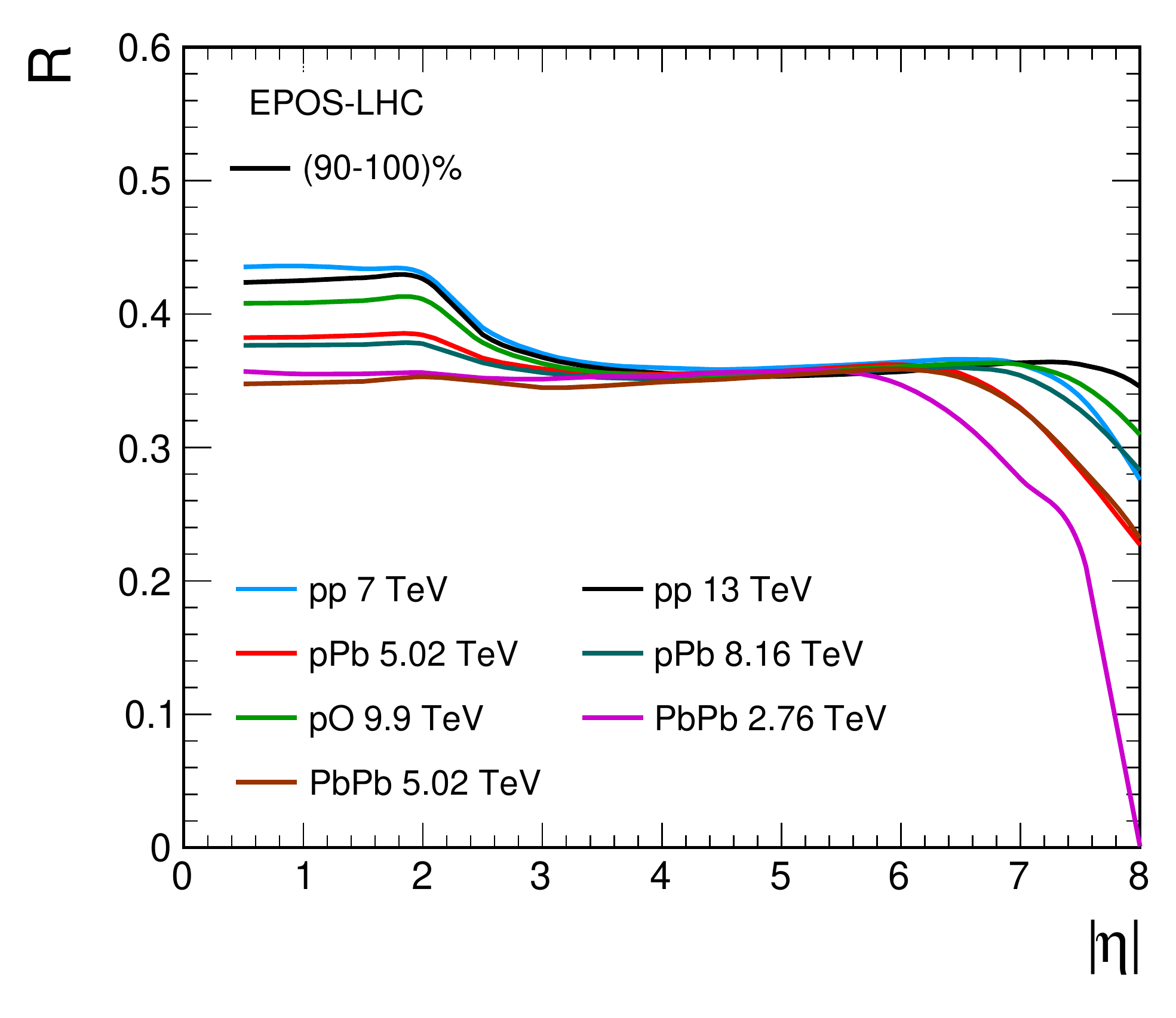}
\caption{(Color Online) Variation of $R$ as a function of pseudorapidity using EPOS LHC over different collision species and energies, considering three multiplicity/centrality classes: (top) High multiplicity (0-10)\%, (middle) Minimum Bias (0-100)\%, and (bottom) Low multiplicity (90-100)\%.}
\label{fig5}
\end{figure}

So far, we have studied strangeness and $R$ at the mid-rapidity region defined by $|\eta|<2$. It has been observed that the particle production mechanisms at the colliders are vastly affected by rapidity, like mid (forward)-rapidity favors gluon (quark) induced processes. A complete description of UCHER-air interactions would require proper modeling of the system over the entire phase space available. Thus, making a rapidity-dependent study of $R$ is vital by slowly increasing the phase space coverage. In Fig.~\ref{fig5}, we perform such a rapidity-dependent study of  $R$ across colliding systems and center of mass energies for minimum biased (MB), most central or high-multiplicity (HM), and peripheral or low multiplicity (LM) data samples generated using EPOS LHC model. A similar trend is followed by all systems for MB and HM cases, although an ordering seems to follow from large to small systems. The value for $R$ is minimum at the mid-rapidity region and slowly increases with increasing phase space coverage before finally falling for $|\eta| \gtrapprox 7$. This could be due to spectators carrying energy mostly towards the forward rapidity region, thus contributing to the hadronic energy fraction. The nature of variation of $R$ in LM or peripheral events, as shown in Fig.~\ref{fig5}, is quite interesting. Peripheral or ultra-peripheral events experience very few hadronic interactions and are governed mainly through electromagnetic interactions, surrounded by relatively large spectators~\cite{Bertulani:2005ru,Dyndal:2017wbv}. 
This can be seen more clearly in the mid-rapidity region ($|\eta|\lessapprox 2$) of small systems due to the relatively less number of baryons involved. The electromagnetic interactions become prominent when the participating nuclei are relatively close, and thus electromagnetic particle production tends to be more prominent at mid-rapidity. In the case of heavier nuclei or larger multiplicities, the mid-rapidity region is dominated by hadrons produced in direct interaction between the nuclei, thus reducing $R$. The higher rapidity regions are, in general, dominated by fragmentation which deposits more energy in the hadronic sector. The effect of fragmentation is the same overall collision species, thus reducing the value of $R$ in small systems and approaching common values. 
These observations put in perspective the importance of the phase space consideration that has to be taken into account for proper modeling of UHECR-air collisions.\\

\begin{figure}[ht!]
\includegraphics[scale = 0.4]{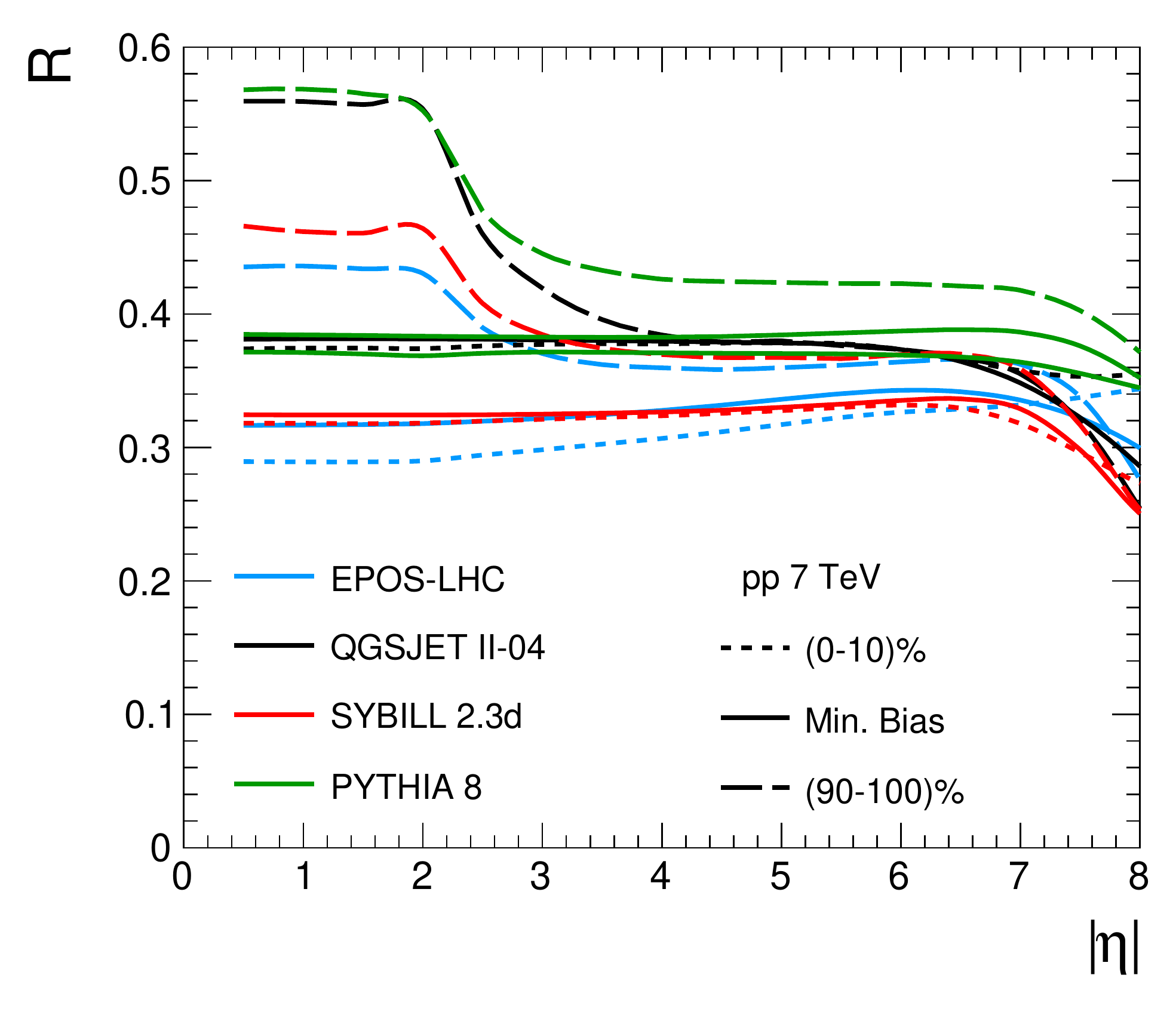}
\caption{(Color Online) Variation of $R$ as a function of pseudorapidity using different models for $pp$ collisions at $\sqrt{s}=$ 7 TeV, considering three multiplicity classes: (i) High multiplicity (0-10)\%, (ii) Minimum Bias (0-100)\%, and (iii) Low multiplicity (90-100)\%.}
\label{fig6}
\end{figure}

In order to check for biases from the model for the observations made from Fig.~\ref{fig5}, we have studied rapidity dependence of the observable $R$ using the different models
for $pp$ collisions at $\sqrt{s}$ = 7 TeV considering MB, HM, and LM events, shown in Fig.\ref{fig6}. For HM and MB event classes, all the models tend to show similar behavior, although $R$ remains almost constant for QGSJET II-04, PYTHIA 8, and SYBILL 2.3d models. In contrast, EPOS LHC shows a variation with increasing pseudorapidity coverage as expected from previous results shown in Fig.~\ref{fig2} and Fig.~\ref{fig3}. For low-multiplicity events, all the models seem to approximately follow a typical behavior with corresponding model variations in the value of $R$. The similarity of the behavior followed by all models suggests that the study/observation of $R$ presented earlier is free of model bias.

\section{Summary}
\label{sum}
This letter focuses on a possible solution to the muon puzzle by considering the possibility of thermalized deconfined medium formation in UHECR-air collisions. This work is inspired by the fact that such collisions might attain very high energy densities, which have been proposed to be explored in the upcoming $p$-O collisions at the LHC. For such collisions to make sense in the context of UHECR-air interactions, new and better observables/parameters are the need of the hour. In this work, we have discussed one such observable called $R$, which could help to understand the energy flow in such collisions. \\

Important findings from this study are summarized below:
 
 \begin{enumerate}
\item Among the four models under consideration, EPOS LHC results match the overall enhancement of strangeness in $pp$, $p$-Pb, and Pb-Pb collisions as a function of multiplicity, though with a more sudden turn-on as compared to data as seen in Fig.\ref{figapp1}. Such an agreement indicates the advantage of the EPOS LHC model at very high energies over other models.

\item Correspondingly, as expected, the energy fraction $R$ decreases with final state-charged particle multiplicity, indicating that correct modeling of strangeness may be achieved through the tuning of $R$.

\item An anticorrelation between strangeness and $R$ is observed in the EPOS LHC model in the $10<  N_{ch}< 200$ range at the mid-rapidity region, which further relates strangeness with the energy fraction $R$ in UHECR-air interactions. 

\item A rapidity-dependent study of $R$ indicates the importance of the complete phase space consideration in determining the energy flow. The HM and HM events show an increase in $R$ towards higher rapidity coverage. In contrast, LM events in small systems show a higher electromagnetic energy fraction in the mid-rapidity region, a trend shown by all the models.

\end{enumerate}

As seen in Ref.~\cite{Schotter:2023khz}, the energy deposited in the zero-degree calorimeter in $pp$ collision can indicate the possibility of strangeness enhancement and medium formation through the anti-correlation of ZDC energy
deposition with the final state charged particle multiplicity and hence the strangeness enhancement. It would thus be worthwhile to look at a similar dependence of $R$ on the energy deposited during the upcoming $p$-O collisions at LHC. This could indicate thermalization in small systems produced in UHECR-air interactions. Such a dependence of $R$ on the energy deposited could further be used to tune the high-energy models, which may help to solve the muon puzzle.

\section*{Acknowledgement}
R. Scaria acknowledges financial support from Council for Scientific and Industrial Research (CSIR), India. S.D. acknowledges the support from the postdoctoral fellowship of CNRS at IJCLAB, Orsay, France. R. Sahoo and C.R.S. acknowledge the financial support under the ALICE project (Project No. SR/MF/PS-02/2021-IITI (E-37123)). The authors acknowledge the Tier-3 computing facility in the experimental high-energy physics laboratory of IIT Indore, supported by the ALICE project.


\section*{APPENDIX}
\appendix
\section{Experimental comparison of Strangeness enhancement}
\label{app_A}
\begin{figure}[ht!]
\includegraphics[scale = 0.4]{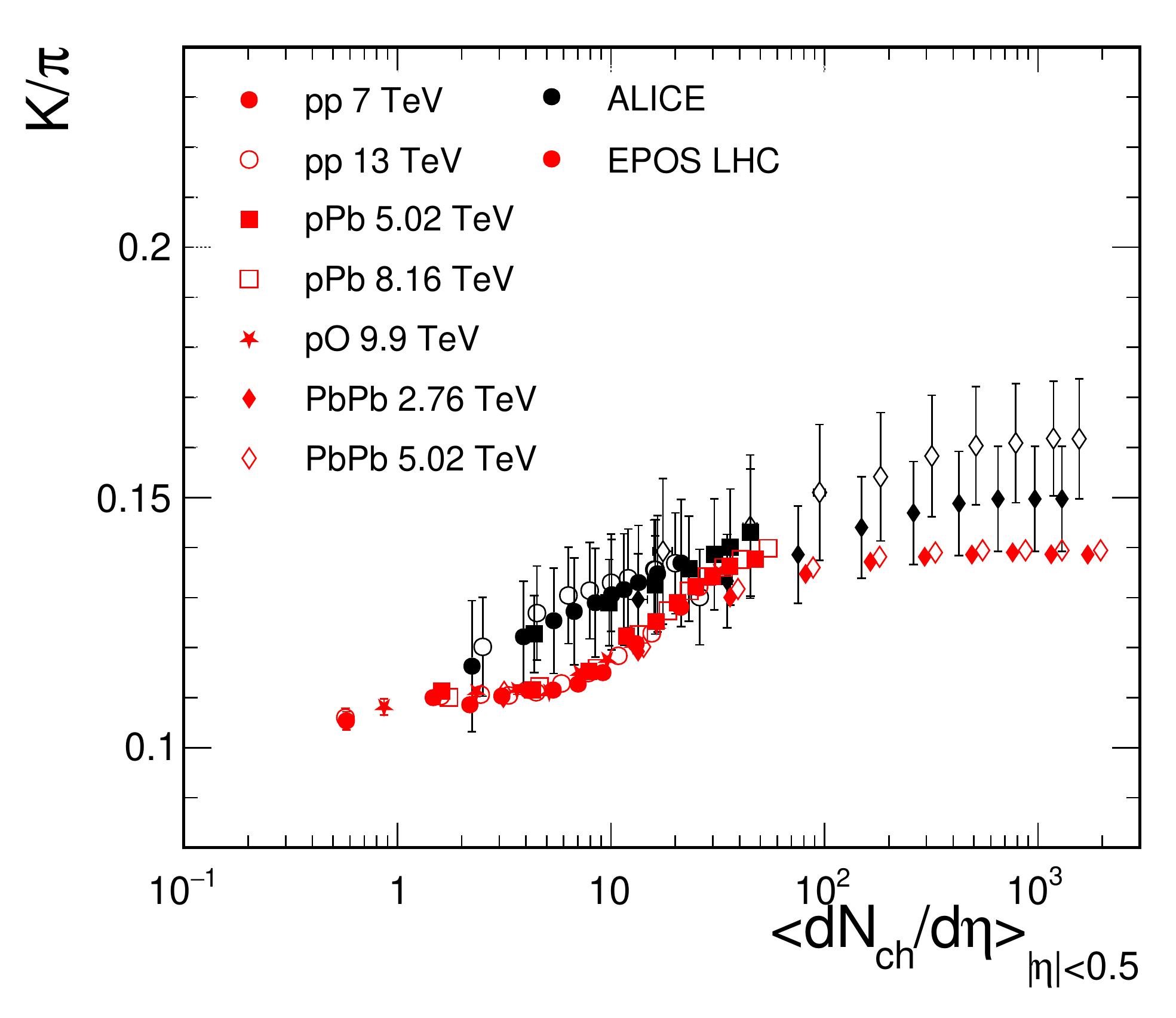}
 \caption{(Color online) Comparison of $K/\pi$ ratio observed at the ALICE experiment \cite{ALICE:2020nkc,ALICE:2018pal,ALICE:2013wgn,ALICE:2013mez,ALICE:2019hno} in the mid rapidity region compared with the results obtained by the EPOS LHC model.}
\label{figapp1}
\end{figure}

The multiplicity dependence of strange particle production ($K/\pi$) observed in the mid-rapidity region at the ALICE experiment \cite{ALICE:2020nkc,ALICE:2018pal,ALICE:2013wgn,ALICE:2013mez,ALICE:2019hno} is shown in Fig.\ref{figapp1}. The results obtained by using the EPOS LHC model are shown for comparison. 


 \end{document}